\documentclass[a4,12pt]{article}
  
\usepackage[margin=2.0cm]{geometry}
\usepackage{graphicx}
\usepackage{amsmath}
\usepackage{rsfso}
\usepackage{breqn}
\usepackage{subfigure}

\begin{document}

\title{
\textbf{Quantum version of Prisoners' Dilemma under Interacting Environment}}

\author{Kaushik Naskar \\ \textit{Department of Physics, Taki Government College}, \\ \textit{West Bengal, India}}

\date{}

\maketitle

\begin{abstract}
Quantum game theory is a rapidly evolving subject that extends beyond physics. In this research work, a schematic picture of quantum game theory has been provided with the help of the famous game Prisoners' Dilemma. It has been considered that the shared qubits of the prisoners may interact with the environment which is a bath of simple harmonic oscillators. This interaction introduces decoherence and this model has been compared with phase damping. Calculation of the decoherence factor shows that decoherence reduces the payoff of the prisoners. The factor has been calculated based on whether the decoherence occurred once or twice. A comparative discussion establishes that the process of decoherence is faster for the case when it occurs twice.  It needs to be mentioned here that the total time of decoherence is the same in both cases.
\end{abstract}

\section{Introduction}
\label{intro}
Game theory has a special aura that enlightens biology, mathematics, social science, and various fields of physics.  The situations where it is difficult to make a decision among multipartite systems, game theory provides a logical approach towards an acceptable solution. Using quantum properties in game theory a new area of physics, quantum game theory has been created, with enormous applications in quantum information and quantum computation. In quantum game theory, the participants can share an entangled quantum state which provides some advantages over the classical game theory. 

\indent Considering two vital properties of quantumness, superposition and entanglement, Eisert  \textit{et al}\cite{Eisert} have introduced the quantum extension of the classical game Prisoners' Dilemma(PD)\cite{Axelrod}. But in the quantum version, the situation changes when the shared qubits of the prisoners interact with the environment. In that case, the state of the environment gets entangled with the qubit and as a consequence quantum decoherence starts ruling the composite state. This decoherence process destroys the quantumness of the system. As it is impossible to construct a perfectly closed system, the interaction of the environment is inevitable. In various literature, the interaction of the system with its environment has been called \textit{monitoring of the environment}. Decoherence reduces the advantages of the players that they gained by the quantum properties. My present work describes the decoherence process and the dependence of payoffs of the prisoners on the time steps for a special model of environmental interaction. The model is also compared with the phase damping channel.

\section{Quantum Game}
\label{QG}

\begin{table}
\centering
\caption{Payoffs for the Prisoners' Dilemma. The numerics in the parenthesis represents the payoffs of the prisoners for chosen strategies C or D. The two digits in a papenthesis denote the payoff of Alice and Bob respectively.}
\label{tab:1}       
\begin{tabular}{lll}
\hline\noalign{\smallskip}
& Bob(C) & Bob(D)  \\
\noalign{\smallskip}\hline\noalign{\smallskip}
Alice(C) & (3,3) & (0,5) \\
Alice(D) & (5,0) & (1,1) \\
\noalign{\smallskip}\hline
\end{tabular}
\end{table}

In the classical Prisoners' Dilemma, two prisoners, Alice(A) and Bob(B) are separately interrogated. They may decide to choose cooperation(strategy C) or defection(strategy D) independently. But their payoffs will depend according to Table \ref{tab:1}. The Nash equilibrium\cite{Nash1}\cite{Nash2} for PD can be found as (D,D). 

\indent In its quantum version\cite{Eisert}, let A and B initially share a 2 qubits state $\left|0 0 \right\rangle$, where $\left|0 \right\rangle$(which denotes C) and $\left|1 \right\rangle$(which denotes D) are 2 basis states of a two dimensional Hilbert space. There is an unitary operator $\hat{J}$ which entangles the two states and known to both of the prisoners. Under the conditions $0\leq \theta \leq \pi$ and $0\leq \phi \leq \pi /2$, two quantum strategies $\hat{U}_A$ (for A) and $\hat{U}_B$ (for B) can be found from,

\begin{equation}
\hat{U}\left(\theta , \phi \right)= \begin{pmatrix}
e^{i\phi} cos(\theta/2) & sin(\theta/2) \\
-sin(\theta/2) & e^{-i\phi} cos(\theta/2)
\end{pmatrix}.
\end{equation}
 Associating the operators for the strategies C and D as $\hat{C}=\hat{U}(0,0)$ and $\hat{D}=\hat{U}(\pi ,0)$ respectively the entanglement operator $\hat{J}$ can be represented by $\hat{J}=\frac{1}{\sqrt{2}}(\hat{I}\otimes \hat{I}+i\hat{D}\otimes \hat{D})$ for maximum entanglement. The general process of a quantum game can be described in the following steps:
\begin{eqnarray}
1.~\text{Initial density operator:~~}~\hat{\rho}_0 &=& \left|00\right\rangle \left\langle 00 \right| \\
2.~\text{Entanglement:~~~~~~~~~~~~~~}~\hat{\rho}_1 &=& \hat{J} \hat{\rho}_0 \hat{J}^\dagger \\
3.~\text{Application of strategies:}~\hat{\rho}_2 &=& \left(\hat{U}_A \otimes \hat{U}_B\right) \hat{\rho}_1 \left(\hat{U}_A \otimes \hat{U}_B\right)^\dagger \\
4.~\text{Final density operator:~~~}~\hat{\rho}_f &=& \hat{J}^\dagger \hat{\rho}_2 \hat{J}
\end{eqnarray}  
After successful execution of the above steps, the prisoners may apply suitable measurements to achieve the payoffs. One of the prisoners(say Alice) may expect the payoff as 
\begin{equation} 
\mathcal{R}_A=\sum_{ij} a_{ij} P_{ij}
\end{equation}
here $P_{ij}$ corresponds to the probability for getting the state $\left|i j \right\rangle$ as a result of the measurement and $a_{ij}$ is the associated classical payoff of Alice. Disappearing the classical Nash equilibrium (D,D) the intervention of quantum physics shows a new Nash equilibrium ($\hat{Q}$,$\hat{Q}$) with higher payoff where $\hat{Q}=\hat{U}(0,\pi/2)$.

\section{Model of Quantum Decoherence}
\label{MD}
\begin{figure}
\subfigure[at $t=0$]{\includegraphics[width=0.5\textwidth]{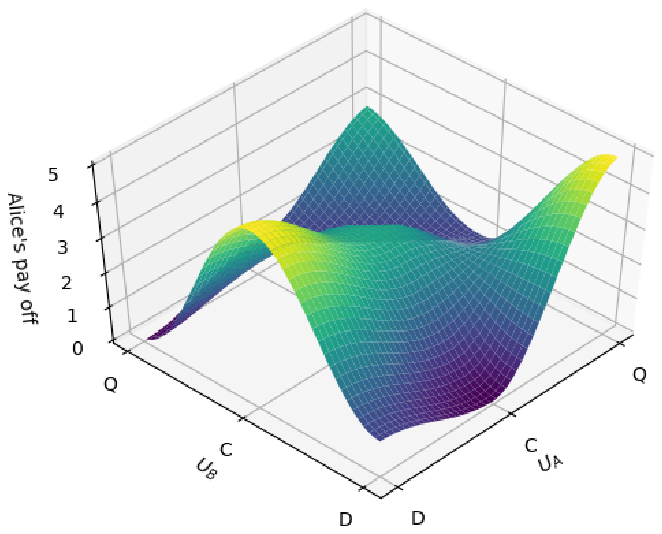}{\label{figa}}}
\subfigure[after $t=1000$]{\includegraphics[width=0.5\textwidth]{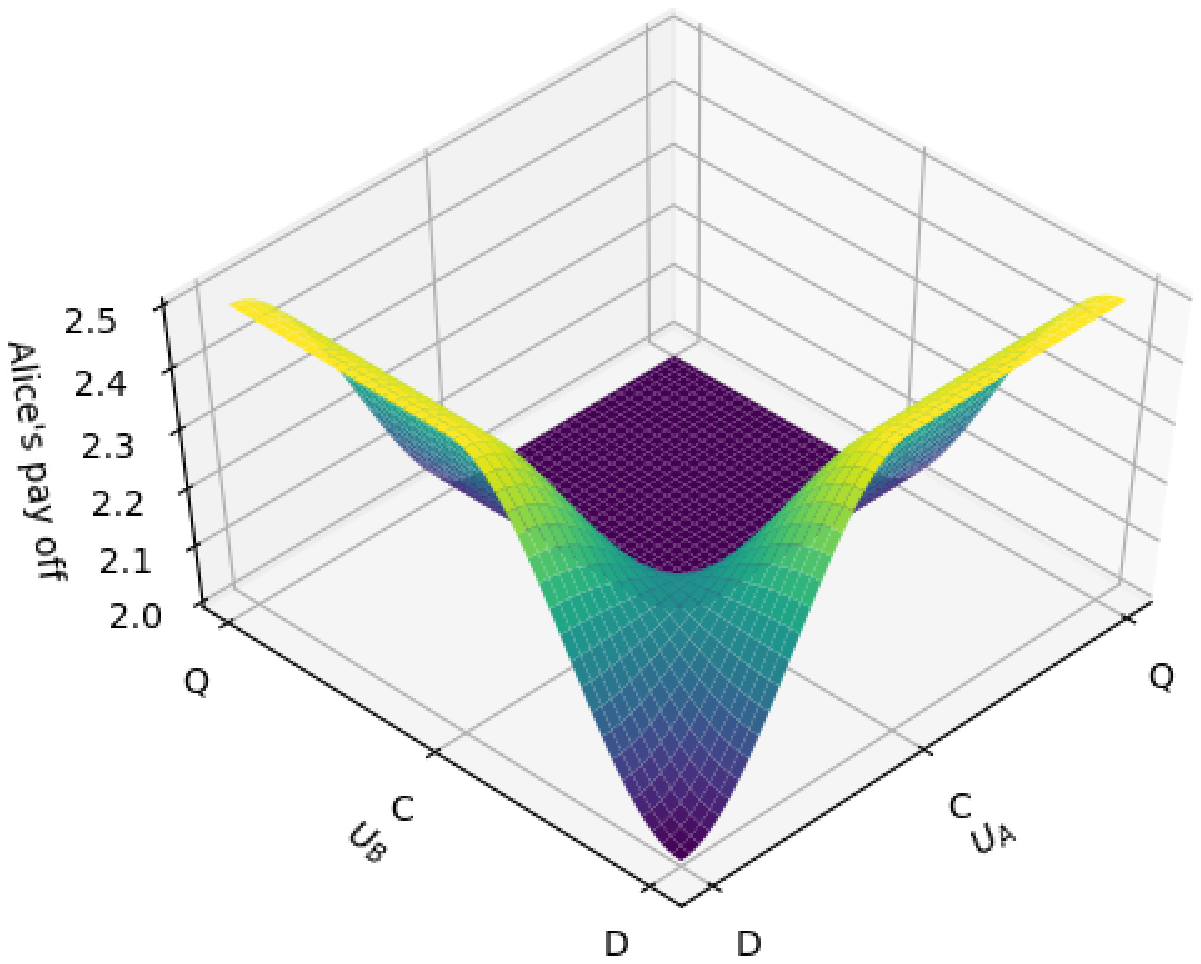}{\label{figb}}}
\caption{Dependence of Alice's payoff on $\hat{U}_A$ and $\hat{U}_B$. It is clear that there is a new Nash equilibrium at $(\hat{Q},\hat{Q})$ at $t=0$. But this Nash equilibrium disappears after $t=1000$. }
\label{Fig1}       
\end{figure}

Along with many advantages, quantum physics also brings some obstacles e.g. quantum decoherence, so that the quality of those advantages decreases. Quantum game theory falls into the trap of quantum decoherence through the environmental monitoring and lost its advantages due to nonunitary evolution. To observe the consequences of quantum decoherence, let us apply a simple process of decoherence between step 2 and step 3 of the general process of quantum game discussed in Section \ref{QG}. Suppose Alice and Bob share an entangled state of two spin$-\frac{1}{2}$ particles and the combined state is interacting with the environment which is a bath of simple harmonic oscillators. The total Hamiltonian$(\hat{H})$\cite{Leggett}\cite{Reina} of the system and the environment takes the form as,
\begin{equation}
\hat{H}=\hat{H}_{A} + \hat{H}_B + \hat{H}_{AE} + \hat{H}_{BE} + \hat{H}_E
\end{equation} 
where $\hat{H}_{A}$, $\hat{H}_B$ and $\hat{H}_E$ denote the Hamiltonian corresponding to Alice, Bob and the environment respectively. $\hat{H}_{AE}$ and $\hat{H}_{BE}$ are symbolized as the interaction Hamiltonian of Alice and Bob with the environment respectively. Explicit expressions of the above Hamiltonians are
\begin{equation}
\hat{H}_{j}=\frac{1}{2}\omega \hat{\sigma}_{zj}, ~~~\hat{H}_{jE}=\hat{\sigma}_{zj} \otimes \sum_i \mathcal{C}_i \hat{x}_i,~~~
\hat{H}_{E}=\sum_i \left( \frac{\hat{p}_i ^2}{2m_i} + \frac{1}{2}m_i \omega_i ^2 \hat{x}_i^2 \right) 
\end{equation}
where $j$ may be treated as $A$ and $B$ for Alice and Bob respectively, $\hat{\sigma}_z$ is one of the well-known Pauli matrices, the Larmour frequency $\omega$ is related to the difference between the energy levels of the two basis states of the associated spin$-\frac{1}{2}$ system, $\mathcal{C}_i$ is the coupling factor between the system and $i^{th}$ harmonic oscillator of the environment and any other terms associated with $i$ like $\hat{p}_i$, $\hat{x}_i$, $m_i$ and $\omega_i$ denote the momentum, position, mass and natural frequency of the $i^{th}$ harmonic oscillator of the environment respectively.

By the help of creation operator$(\hat{a}^\dagger)$ and annihilation operator$(\hat{a})$ the Hamiltonian and the time evolution operator in the interaction picture may be written as
\begin{eqnarray}
\hat{H}_I (t)&=&(\hat{\sigma}_{zA}+\hat{\sigma}_{zb})\otimes \sum_i \left(\mathcal{G}_i \hat{a}^\dagger _i e^{i\omega_i t} + \mathcal{G}^*_i \hat{a}_i e^{-i \omega_i t}\right), \\
\hat{U}_I (t)&=&\mathcal{P} e^{\frac{1}{2}(\hat{\sigma}_{zA}+\hat{\sigma}_{zB})\otimes\sum_i \left(\eta_i (t)\hat{a}^\dagger _i -\eta^* _i (t)\hat{a}_i\right)}
\end{eqnarray}
respectively, where $\mathcal{G}_i$ is considered as complex for generic case and related to $\mathcal{C}_i$ as $\mathcal{C}_i \hat{x}_i =(\mathcal{G}_i \hat{a}^\dagger _i + \mathcal{G}^* _i \hat{a}_i)$, $\eta_i (t)$ can be found as $\eta_i (t)=2 \frac{\mathcal{G}_{i}}{\omega_i}\left(1-e^{i\omega_i t}\right)$ and $\mathcal{P}$ is a global phase term which has no importance for our case and may be neglected.

Now let us suppose that after application of the entanglement operator $\hat{J}$ the combined state of Alice and Bob becomes $|\psi (0)\rangle=\frac{1}{\sqrt{2}}\left[|00\rangle +i|11\rangle \right]$ which is maximally entangled. If the environment is in thermal equilibrium at temperature $T$ then the combined density operator for Alice, Bod and the environment will be,
\begin{equation*}
\hat{\rho}_{ABE} (0)=\frac{1}{2}\left[|00\rangle \langle 00| -i|00\rangle \langle 11| +i|11\rangle \langle 00| +|11\rangle \langle 11|\right] \otimes_i \frac{e^{-\beta \omega_i \hat{a}^\dagger _i \hat{a}_i}}{Tr(~e^{-\beta \omega_i \hat{a}^\dagger _i \hat{a}_i})}
\end{equation*}
where $\beta=1/K_B T$ and $K_B$ is Boltzmann constant. After certain time $t$, the total density operator will be evolved to $\hat{\rho}_{ABE}(t)=\hat{U}_I (t)\hat{\rho}_{ABE} (t) \hat{U}^{-1} _I (t)$ and the reduced density operator for Alice and Bob will be $\hat{\rho}_{AB}(t)=Tr_E \left[\hat{\rho}_{ABE}(t)\right]$. Calculations show that the diagonal elements of $\hat{\rho}_{AB}(t)$ remain constant. But in case of the off-diagonal elements the situation is different. The off-diagonal elements of $\hat{\rho}_{AB}(t)$ are
\begin{equation}
\left[\hat{\rho}_{AB}(t)\right]_{kl}=\left[\hat{\rho}_{AB}(0)\right]_{kl} \prod_i \left\langle e^{\pm 2\left[\eta_i (t)\hat{a}^\dagger _i -\eta^* _i (t) \hat{a}_i\right]}\right\rangle_E
\end{equation}
where $k$ and $l$ can have values $0$ or $1$ but $k\ne l$. In various literature\cite{Qoptics}\cite{Open} it can be found that,
\begin{equation}
\left\langle e^{ 2\left[\eta_i (t)\hat{a}^\dagger _i -\eta^* _i (t) \hat{a}_i\right]}\right\rangle_E = e^{-|\eta_i (t)|^2 coth(\frac{\beta \omega_i}{2})}
\end{equation}
and the term,
\begin{equation}
\prod_i e^{-|\eta_i (t)|^2 coth(\frac{\beta \omega_i}{2})}=\mathcal{D}(t)
\label{DF}
\end{equation}
is responsible for the time dependence of the off-diagonal terms of the reduced density operator and is called decoherence factor or decoherence function. 

For a large and dense environment, the environmental frequency modes can be considered as continuous. If we use Ohmic spectral density with a high-cutoff frequency $\omega_c$, then eqn.(\ref{DF}) takes the form as
\begin{equation}
\mathcal{D}(t)=exp\left[-J_0 \int_0 ^\infty \frac{d\omega}{\omega} coth(\frac{\beta \omega}{2}) (1-cos~\omega t) exp(-\frac{\omega}{\omega_c})\right]
\end{equation}
where $J_0$ is a constant. In the next section, the dependence of payoff on the decoherence factor $\mathcal{D}(t)$ will be discussed.

\section{Result and Analysis}
\label{RA}
\begin{figure}
\centering
\includegraphics[width=0.5\textwidth]{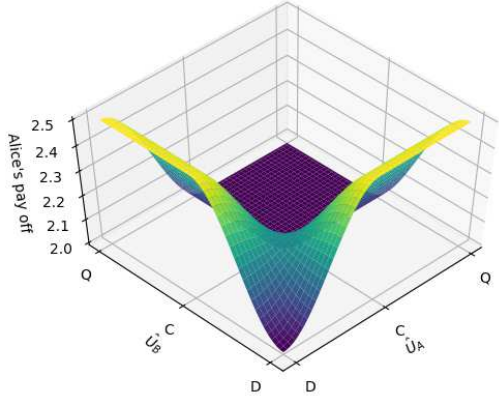}
\caption{Dependence of Alice's payoff on $\hat{U}_A$ and $\hat{U}_B$ after the total decoherence time $t=1000$ when decoherence takes place before the application of the strategies and after the application of their strategies. This shows a similar effect as in Figure \ref{figb}.}
\label{Fig2}       
\end{figure}

After decoherence, if the prisoners apply the quantum strategies $\hat{U}_A$ and $\hat{U}_B$ to their respective quantum states then one of the prisoners(say Alice) can have the payoff as 
\begin{dmath}
\mathcal{R}_A = [2+\mathcal{D}(t)cos\{2(\phi_A+\phi_B)\}]cos^2 \left(\frac{\theta_A}{2}\right) cos^2\left(\frac{\theta_B}{2}\right) + \{2-\mathcal{D}(t)\}sin^2\left(\frac{\theta_A}{2}\right)sin^2\left(\frac{\theta_B}{2}\right)- \left[5 sin(\phi_A -\phi_B)+sin(\phi_A +\phi_B)\{\mathcal{D}(t)+2\}\right]
sin\left(\frac{\theta_A}{2}\right)sin\left(\frac{\theta_B}{2}\right)cos\left(\frac{\theta_A}{2}\right)cos\left(\frac{\theta_B}{2}\right)
+ \frac{5}{2}\left[\{1-\mathcal{D}(t)cos(2\phi_A)\}cos^2\left(\frac{\theta_A}{2}\right)sin^2\left(\frac{\theta_B}{2}\right)+\{1+\mathcal{D}(t)cos(2\phi_B)\}sin^2\left(\frac{\theta_A}{2}\right)cos^2\left(\frac{\theta_B}{2}\right) \right].
\label{Deco1}
\end{dmath}  

Considering the condition $\omega <\omega_c$ and using some numerical approach for $\mathcal{D}(t)$, equation (\ref{Deco1}) has been represented pictorially by Figure \ref{Fig1}. Figure \ref{figa} shows the distribution of Alice's payoff at $t=0$, which is similar to Eisert \textit{et al}\cite{Eisert} and that after $t=1000$ is shown by Figure \ref{figb}, which is consistent with Huang and Qiu\cite{Huang}. The Nash equilibrium ($\hat{Q}$,$\hat{Q}$) disappears as the time step runs from 0 to 1000. These results signify that the model of quantum decoherence discussed in Section \ref{MD} is a model for phase damping applied to quantum game theory.

Now suppose the entangled quantum states of Alice and Bod interact with the same environment for the second time between step 3 and step 4 of Section \ref{QG}. Then again we need to perform the tedious job of Section \ref{MD} and finally the payoff of Alice will be obtained as,
\begin{figure}
\centering
\includegraphics[width=0.5\textwidth]{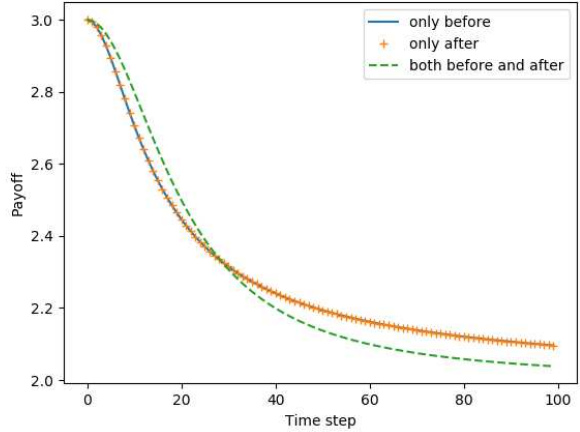}
\caption{Dependence of Alice's payoff on decoherence time step when both of the prisoners apply $\hat{Q}$ to their respective quantum states. It shows that the payoff reduces faster when the decoherence occurs twice, i.e., before the application of the strategies and after the application of the strategies.}
\label{Fig3}       
\end{figure}

\begin{dmath}
\mathcal{R}_A=cos^2\left(\frac{\theta_A}{2}\right)cos^2\left(\frac{\theta_B}{2}\right) \{2+\mathcal{D}_1 (t_1) \mathcal{D}_2 (t_2) cos(2\phi_A +2\phi_B)\} + sin^2\left(\frac{\theta_A}{2}\right)sin^2\left(\frac{\theta_B}{2}\right)\{2-\mathcal{D}_1 (t_1) \mathcal{D}_2 (t_2) \} -sin\left(\frac{\theta_A}{2}\right)sin\left(\frac{\theta_B}{2}\right) cos\left(\frac{\theta_A}{2}\right)cos\left(\frac{\theta_B}{2}\right)\{sin(\phi_A +\phi_B)\left(\mathcal{D}_1(t_1)+2\mathcal{D}_2 (t_2)\right)+5 sin(\phi_A -\phi_B)\mathcal{D}_2 (t_2)\} + \frac{5}{2}\left[cos^2 \left(\frac{\theta_A}{2}\right) sin^2 \left(\frac{\theta_B}{2}\right)\{1-\mathcal{D}_1 (t_1) \mathcal{D}_2 (t_2) cos(2\phi_A)\}+sin^2 \left(\frac{\theta_A}{2}\right) cos^2 \left(\frac{\theta_B}{2}\right)\{1+\mathcal{D}_1 (t_1) \mathcal{D}_2 (t_2) cos(2\phi_B)\}\right]
\label{finalDeco}
\end{dmath} 
where $\mathcal{D}_1 (t_1)$ and $\mathcal{D}_2 (t_2)$ are the decoherence factors for the decoherence processes that occurred before and after the application of the strategies respectively and $t_1$ and $t_2$ are the respective time elapsed during the decoherence processes. Figure \ref{Fig2} depicts the situation of equation (\ref{finalDeco}) after the total decoherence time $t=1000$, where $t=t_1 +t_2$. Both Figure \ref{figb} and Figure \ref{Fig2} show that the dependence of Alice's payoff on the strategies $\hat{U}_A$ and $\hat{U}_B$ is the same after large numbers of time steps like $t=1000$. Figure \ref{Fig3} represents the time dependence of Alice's payoff when both of the prisoners apply $\hat{Q}$ to their respective qubits. It reflects two different situations depending upon the occurrence of the decoherence: (i) when decoherence occurs only before the application of the strategies or only after the application of the strategies, the payoff follows the same curve but (ii) the total effect of decoherence is faster when it takes place both before the application of the strategies and after the application of the strategies than in the previous cases though the total time of decoherence is equal for all the cases. 

Since no ideally closed system is possible monitoring of the environment always affects the quantumness of a quantum system through decoherence. In this work, presenting a model for quantum decoherence and applying the model to the quantum version of Prisoners' Dilemma it has been shown that the advantages of quantum game theory disappear with time. This model of decoherence is compared with phase damping. The dependence of payoff on time steps has also been discussed in different cases.   

\section{Acknowledgements}
I would like to thank Dr. Sankhasubhra Nag, Department of Physics, Ramakrishna Mission Vivekananda Centenary College, Rahara, Kolkata-118, for many fruitful discussions on Quantum Decoherence. \\
I am grateful to Dr. Parthasarathi Joarder, Department of Physics, Sister Nivedita University, New Town, Kolkata-156, for continuous worthwhile academic discussions.

\end{document}